\documentclass{emulateapj}

\usepackage[ansinew]{inputenc}
\usepackage{hyperref}
\usepackage{amsmath}
\usepackage{color}

\def\mms{$\mathrm{mm\,s^{-1}}$}
\def\cms{$\mathrm{cm\,s^{-1}}$}
\def\metersecond{$\mathrm{m\,s^{-1}}$}

\shorttitle{Low-velocity collisions of dust aggregates}
\shortauthors{Beitz et al.}
\slugcomment{Accepted by The Astrophysical Journal}

\begin{document}

\title{Low-velocity collisions of centimeter-sized dust aggregates}

\author{E. Beitz, C. Güttler, J. Blum}
\affil{Institut für Geophysik und extraterrestrische Physik, Technische Universität zu Braunschweig}
\email{e.beitz@tu-braunschweig.de}

\and

\author{T. Meisner, J. Teiser, G. Wurm}
\affil{Fakultät für Physik, Universität Duisburg-Essen}

\begin{abstract}
Collisions between centimeter- to decimeter-sized dusty bodies are important to understand the mechanisms leading to the formation of planetesimals. We thus performed laboratory experiments to study the collisional behavior of dust aggregates in this size range at velocities below and around the fragmentation threshold. We developed two independent experimental setups with the same goal to study the effects of bouncing, fragmentation, and mass transfer in free particle-particle collisions. The first setup is an evacuated drop tower with a free-fall height of 1.5 m, providing us with 0.56 s of microgravity time so that we observed collisions with velocities between 8 \mms\ and 2 \metersecond. The second setup is designed to study the effect of partial fragmentation (when only one of the two aggregates is destroyed) and mass transfer in more detail. It allows for the measurement of the accretion efficiency as the samples are safely recovered after the encounter. Our results are that for very low velocities we found bouncing as could be expected while the fragmentation velocity of 20 \cms\ was significantly lower than expected. We present the critical energy for disruptive collisions $Q^{\star}$, which showed up to be at least two orders of magnitude lower than previous experiments in the literature. In the wide range between bouncing and disruptive collisions, only one of the samples fragmented in the encounter while the other gained mass. The accretion efficiency in the order of a few percent of the particle's mass is depending on the impact velocity and the sample porosity. Our results will have consequences for dust evolution models in protoplanetary disks as well as for the strength of large, porous planetesimal bodies.
\end{abstract}


\section{Introduction}
In a protoplanetary disk (PPD) a large number of collisions between dust agglomerates takes place. The relative velocities between these particles, caused by Brownian motion, differential drift motions, and gas turbulence are depending on their sizes \citep[see, e.g., ]{WeidenschillingCuzzi:1993,WeidlingEtal:2009}. According to the simulations of \citet{ZsomEtal:2010a} the largest particles growing in a minimum mass solar nebula (MMSN) model are about two centimeters in diameter. For particles of this size, collision velocities of about 10 \cms\ are expected. However, different PPD models predict a wide range of collision velocities for cm-sized dust agglomerates, ranging from about 1 \cms\ for the model of \citet{Desch:2007} to about 5 \metersecond\ for the low-density model by \citet{AndrewsWilliams:2007}. Thus, it is interesting to experimentally investigate the collision behavior of cm-sized dust aggregates in a wide range of velocities also to assess the validity of the dust-collision model by \citet{GuettlerEtal:2010} and the dust-evolution model by \citet{ZsomEtal:2010a}. For low-velocity collisions between cm-sized dust aggregates, it is indispensable to design an experimental setup in which the collisions take place without the overwhelming influence of gravity. In Sect. \ref{sec:experimental_setup}, we describe two experimental setups with which we could investigate free collisions between two cm-sized dust aggregates in a wide velocity regime between 8 \mms\ and 2 \metersecond. Moreover, the choice of our analog dust particles is explained. Sect. \ref{sec:results} shows the results of these experiments and Sect. \ref{sec:discussion} discusses and interprets these results in terms of their applicability in PPD dust-evolution models.

\section{Experimental Setup}\label{sec:experimental_setup}
To study the effects of dust aggregate collisions for a variety of collision velocities, two slightly different experimental setups were developed: in the drop-tower setup (Sect. \ref{sec:experimental_setup_1}) we were able to observe free collisions between cm-sized dust aggregates for very low to intermediate impact velocities (0.8--200~\cms). In a second setup, in which we studied velocities above the fragmentation threshold (approx. 50~\cms, Sect. \ref{sec:experimental_setup_2}), the samples were -- in contrast to the drop-tower setup -- not demolished after the encounter, so that a detailed measurement of the mass balance was possible. We used this setup to study the mass transfer between the aggregates as a function of the porosity difference and the collision velocity.

\begin{figure}[!t]
    \includegraphics[width=9cm]{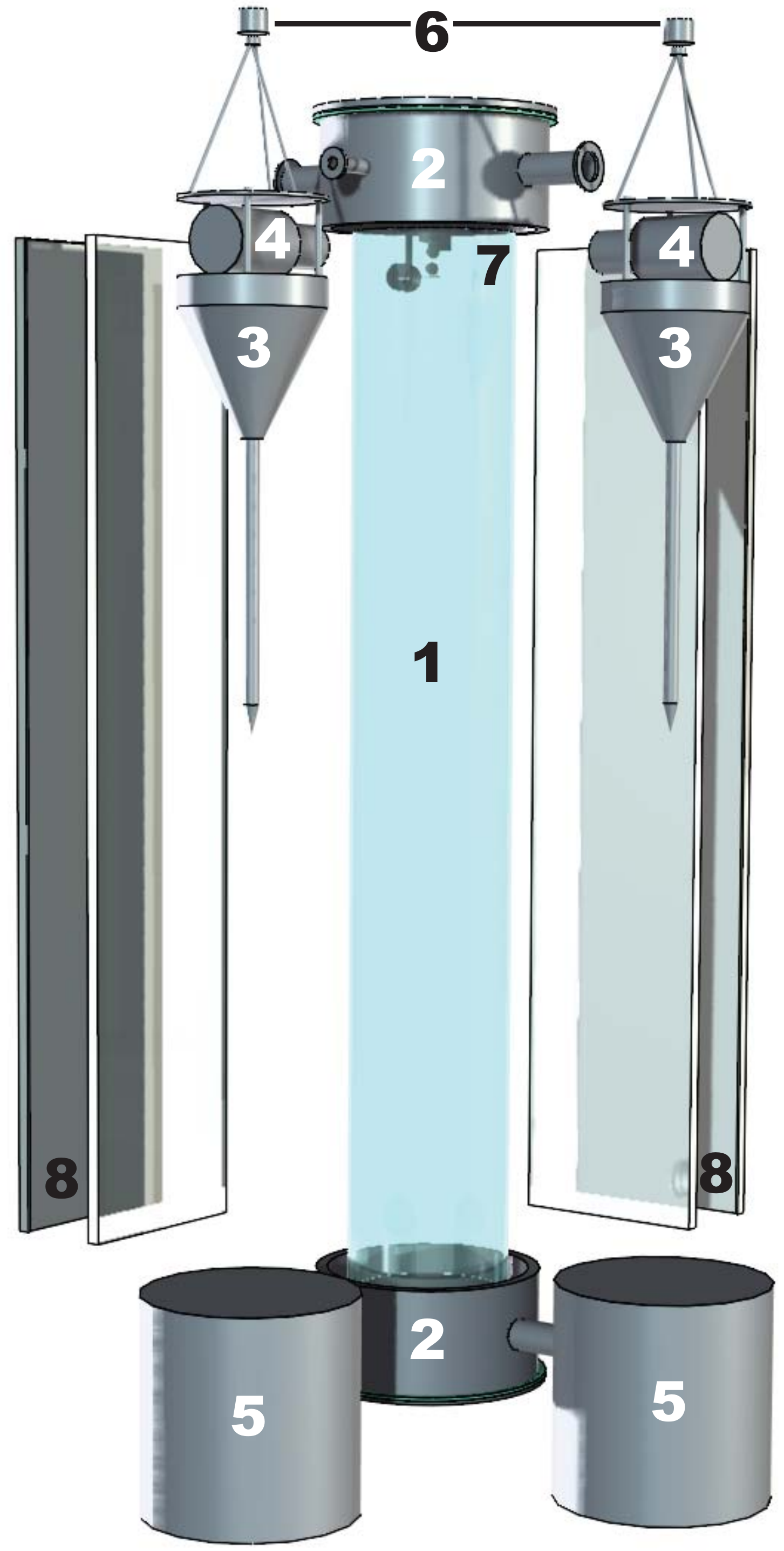}
    \caption{\label{fig:tower}The drop tower setup consists of \textit{(1)} a glass tube with a height of 1.5 m and an inside diameter of 22 cm, \textit{(2)} vacuum cambers with electrical feedthroughs, pressure gauge and vacuum pump, \textit{(3)} camera platforms with a streamline funnel, \textit{(4)} two high-speed cameras separated by an angle of $90^{\circ}$, \textit{(5)} buckets filled with sand to decelerate the cameras, \textit{(6)} electromagnets for attaching the camera platforms to the ceiling, \textit{(7)} a two-level particle release mechanism (also see Fig. \ref{fig:release-mechanism}), and \textit{(8)} two LED arrays and diffusion screens for back-light illumination.}
\end{figure}

\subsection{Drop-Tower Setup}\label{sec:experimental_setup_1}
The first setup is an evacuated drop tower consisting of a glass tube with two vacuum chambers attached to the top and to the bottom. Outside the glass tube, two cameras are recording the experiment in back-light illumination (Fig. \ref{fig:tower}).

\begin{figure}[t]
    \includegraphics[width=\columnwidth]{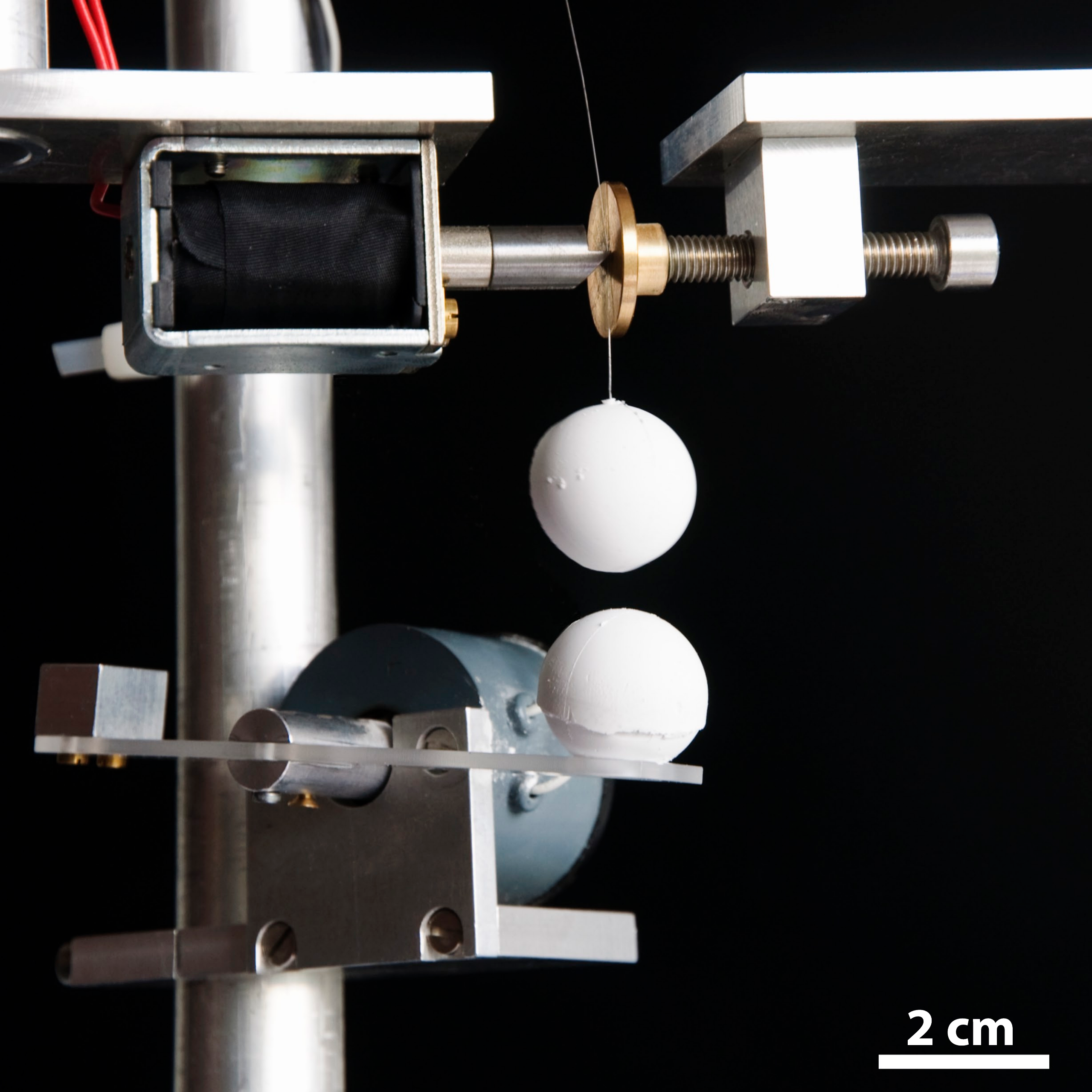}
    \caption{\label{fig:release-mechanism}The two-level release-mechanism: the upper level consists of a linear solenoid, pressing a nylon string with a metal rod against an adjustable metal plate. The lower level is a plate with a hole mounted on a rotary solenoid.}
\end{figure}

The main part of the drop tower is the glass tube with an height of 1.5 m and 22 cm inside diameter. Two vacuum chambers are fixed at the top and at the bottom of the glass tube, respectively. Electrical feedthroughs are attached to these chambers to provide the power for a magnetic release mechanism (see below). Additionally, a pressure gauge monitors the vacuum quality, which was $1-10$~Pa in all experiments. To perform collisions between two 2~cm diameter dust aggregates, a two-level release-mechanism (Fig. \ref{fig:release-mechanism}) is attached at the top flange of the upper chamber. The upper level consists of a linear solenoid pushing a nylon string embedded in the sample onto a vertically mounted plate. The lower level comprises a plate with a hole, mounted to a rotary solenoid, where the particle size exceeds the hole diameter and is thus centered in position. By moving the position of the vertical plate, the impact parameter can be adjusted precisely. The distance between the upper and lower particle edges can be varied between 1~mm and 30~cm.

To be able to record a collision in free fall, two high-speed cameras are dropped outside the glass tube with their fields-of-view centered at the center of mass of the free-falling particles. The cameras are each placed on a platform mounted on the top of a streamlined funnel that ends up with a steel rod. Due to their shape and weight of about 10~kg, they are trailing by less than 5 mm versus the particles (which fall in vacuum) at the end of their free fall. Prior to the experiment run, the camera platforms are fixed at the ceiling with an electromagnet that can be released by switching off the current. The two cameras are separated by an angle of $90^{\circ}$, which provides a three dimensional observation of the collision. One camera is operated at a picture rate of 261 frames per second with a field-of-view of about $\mathrm{6 \times 6\, cm^2}$ and a resolution of 190 pixels $\rm cm^{-1}$. The other one is a photo camera, recording with 40 frames per second and a significantly larger field-of-view ($\mathrm{15 \times 15\, cm^2}$) at a comparable resolution. A back-light illumination was realized by two LED arrays with diffusion screens at the opposite side of the glass tube.

The experiment is triggered  by a microcontroller which controls two time periods: the first one is the relative delay between the release of the upper level and the drop of the falling cameras. The second one is the time lag between the cameras and the lower level. The reachable minimal sum of the two time lags is in the order of one millisecond, which leads to the smallest achievable collision velocity of
\begin{equation}
    v_{\mathrm coll} = g\cdot \Delta t=9.81\,\mathrm{m\,s^{-2}}\cdot 1\,\mathrm{ms}\approx 1\,\mathrm{cm\,s^{-1}},
\end{equation}
where $g$ is the gravitational acceleration. The upper limitation of the accomplishable velocities is based on the maximal distance between the edges of the particles of about $\Delta h = 0.3$~m, which yields a collision velocity of $v_{\mathrm coll}=\sqrt{2g\Delta h}\approx2.5$~\metersecond.

It is desirable to adjust the collision time to one half of the free-fall time, so that the same number of images is available for data analysis before and after the collision. For this, the distance between the particles $\Delta h$ must be adjusted according to the equation
\begin{equation}
    \Delta h = v_\mathrm{coll}\left(\frac{t_{\mathrm{freefall}}}{2}-\frac{v}{2g}\right)
\end{equation}
with a free fall time for the 1.5~m long drop tower which is $t_{\mathrm{freefall}} = 0.55$~s.

The samples used for the particle-particle collisions in this experiment are manufactured in a self-built compression mechanism. It consists of a mold from stainless steel and an aluminum stamp with a concave shape at one end. The stamp fits into the mold such that a ball of 2~cm diameter can be compressed inside. To avoid the dust sticking to the mechanism, the surface was worked with a nanotechnics polish. For half of the spheres, a nylon string is embedded, on which they were suspended in the upper level of the particle release mechanism (cf. Fig. \ref{fig:release-mechanism}).

For each sample, we poured exactly 4.1 g of spherical SiO$_2$ dust (see Sect. \ref{sec:analog_material} for details) into the container and compressed it with the stamp. As this pressure could not be perfectly identical in all cases, we expect small variation in the volume filling factors between $\phi = 0.49$ and $\phi = 0.54$. These values were inferred from the known mass and the volume measurement of the falling aggregates for those cases where the spheres were fully visible. Thus, it is not expected that two colliding aggregates are perfectly similar but one will always be slightly more porous and fragile than the other, which will become important in the discussion. The average difference in volume filling factor between two colliding spheres is $\Delta\phi = 0.025 \pm 0.025$.

\subsection{Solenoid-Accelerator Setup}\label{sec:experimental_setup_2}
In this series of experiments, collisions between dust aggregates with different filling factors at different velocities were studied in a velocity range from 0.5 to 1.4~\metersecond\ and a range in porosity differences from 0.029 to 0.129. Dust aggregates of irregular SiO$_2$ dust (see Sect. \ref{sec:analog_material}) were prepared in a similar fashion as for the drop-tower experiments (Sect. \ref{sec:experimental_setup_1}) by compressing dust into a given volume, while we did not aim to produce spherical particles here but rather compressed cylinders of 30 mm diameter and variable heights. We always used the same mass of $(13.02 \pm 0.25)$~g and by qualitatively varying the compressing pressure we generated dust aggregates, which varied in filling factor between $\phi = 0.35$ and $\phi = 0.49$. This changes the thickness of the cylindrical aggregate slightly but typically the cylindrical dust aggregate had a height of about 17~mm. In the drop-tower experiments the aggregates are destroyed after the collision because they simply crash on the bottom of the experiment chamber. Complementary to this, we aimed here to recover the aggregates after the experiment to allow mass measurement, i.e. to determine the mass balance of the aggregate.
\begin{figure}[t]
    \includegraphics[width=\columnwidth]{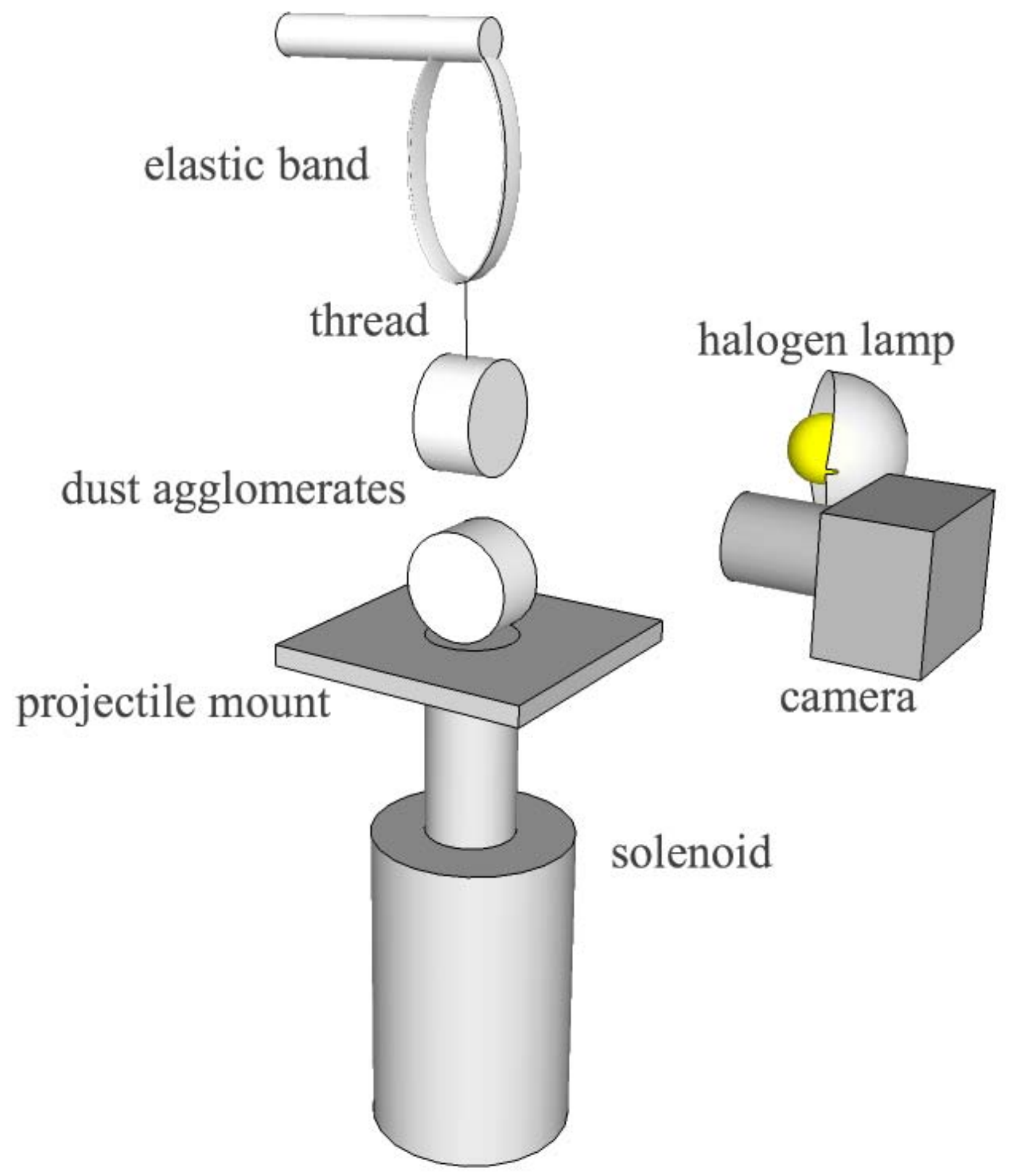}
    \caption{\label{fig:sole}The experimental setup for collisions of the dust cylinders. The solenoid accelerates the lower aggregate, which collides with the upper aggregate which is suspended un an elastic strip.}
\end{figure}
To accomplish this, one of the dust aggregates was suspended on a thread which itself was fixed to an elastic strap. The second agglomerate was launched vertically upwards and hit the target. The target obtained an upward momentum and was somewhat lifted while the elastic band decelerated it slow enough on its downward motion so that it remained intact and any gained mass stayed on the aggregate. With this setup the particle could behave in the same way as in a free collision and the gained mass stayed intact on the target so this could be weighted precisely after the encounter.\ From first experiments it turned out that the more compact aggregate (higher filling factor) survived the collision and gained mass from the less compact aggregate, while this fragmented in the collision. Therefore, the compact aggregate was suspended and the less compact aggregate was accelerated. No alteration of the aggregates' conditions could be observed due to the embedding of the string.\ As stable configuration the compact dust aggregate always had the same filling factor of $0.480 \pm 0.006$. The setup is sketched in Fig. \ref{fig:sole}. The launcher consists of a solenoid, on which a small platform is mounted. On this platform the dust aggregate with higher porosity (i.e. the weaker aggregate) is placed. We observed central collisions with aggregates colliding with their mantle face.

The whole assembly was mounted in a vacuum chamber with an ambient pressure of 0.1~Pa to avoid gas effects during the collision. The collisions at the height of the target were captured in an image sequence with 2 ms time difference between the frames. Based on these movies, collision velocities were determined and it was assured that the mass gained in a collision remained on the target during deceleration. The mass of the target was always measured before and after the collisions.

\subsection{Choice of Analog Material}\label{sec:analog_material}
To prepare samples for an experiment that aims to study the formation of planets, proper analog material has to be chosen. An additional requirement is the commercial availability of the material. We therefore used SiO$_2$ dust of different shapes in the micrometer-size range. This dust has been used in many previous experiments \citep{BlumWurm:2008}, which showed that it can -- at least in a mechanical sense -- be regarded as representative for the class of silicates. For the experiments with spherical aggregates in the drop-tower experiments we used monomers of 1.5~$\mu$m diameter (manufacturer: micromod). These grains have been well characterized by \citet{HeimEtal:1999} and also the material properties were analyzed by \citet{BlumSchraepler:2004} and \citet{GuettlerEtal:2009}. For the cylindrical samples in the second setup we used irregular SiO$_2$ grains in a size range from 0.1 to 10~$\mu$m where 80~\% of the mass are in the range between 1 and 5~$\mu$m (manufacturer: Sigma-Aldrich). Those irregular grains have also been used in many previous experiments \citep{BlumWurm:2008}. From these experiments we learned that the grain shape plays a minor role for the aggregate behavior and those two species can carefully be regarded as comparable. We will discuss potential differences in Sect. \ref{sec:discussion}.

\section{Results}\label{sec:results}
In this section, we will present the results of our experiments, which includes the measurement of coefficients of restitution, strength of fragmentation and mass transfer. In Sect. \ref{sec:BC} we analyze 24 collisions between spherical dust aggregates in a velocity rage of \mbox{0.8--37~\cms}, which all led to bouncing. For the higher collision velocities, we present eleven collisions of spherical aggregates from 24 to 202~\cms, which are described in Sect. \ref{sec:PFC}. In those collisions, one or both aggregates fragmented, and in the cases where only one aggregate fragmented, the preserved aggregate even gained mass. Above 190~\cms, all collisions led to the disruption of both aggregates. The accretion of one aggregate at the intermediate collision velocity is studied in further detail in Sect. \ref{sec:R_ShootingSetup}. There, we will present results from collisions between cylindrical dust aggregates, in which we varied not only the collision velocity but also the porosity difference between the aggregates. We performed 24 experiments at a constant collision velocity ($v=93$~\cms) and varied the porosity difference between projectile and target, then 20 experiments with a constant porosity difference ($\Delta\phi=0.091$) and variable velocities. The particle properties and experimental conditions are shown in the table \ref{table1}.

\begin{table*}
\center

\caption{Overview on the conducted experiments and their outcomes}
\label{table1}
\footnotesize
\begin{tabular}{lccccccl}
  \hline \hline
  experiment type & repetitions & pressure [Pa]&$\mathrm{v_{coll}} $[\cms]& $\phi_1$& $\phi_2$& mass [g]&collisional outcome\\
  \hline
  drop tower           &24 &1--10  & 0.8--37 & 0.5        & 0.5  & 4.1 & bouncing\\
  drop tower           &6  & 1--10 & 47--186 & 0.5        & 0.5  & 4.1 & partial fragmentation\\
  drop tower           &5  & 1--10 & 24--202 & 0.5        & 0.5  & 4.1 & fragmentation\\
  solenoid accelerator &20 & 0.1   & 50--140 & 0.39       & 0.48 & 13  & fragmentation\\
  solenoid accelerator &24 & 0.1   & 93      & 0.34--0.45 & 0.48 & 13  & fragmentation\\
  \hline
\end{tabular}
\end{table*}

\begin{figure}[t]
    \includegraphics[width=\columnwidth]{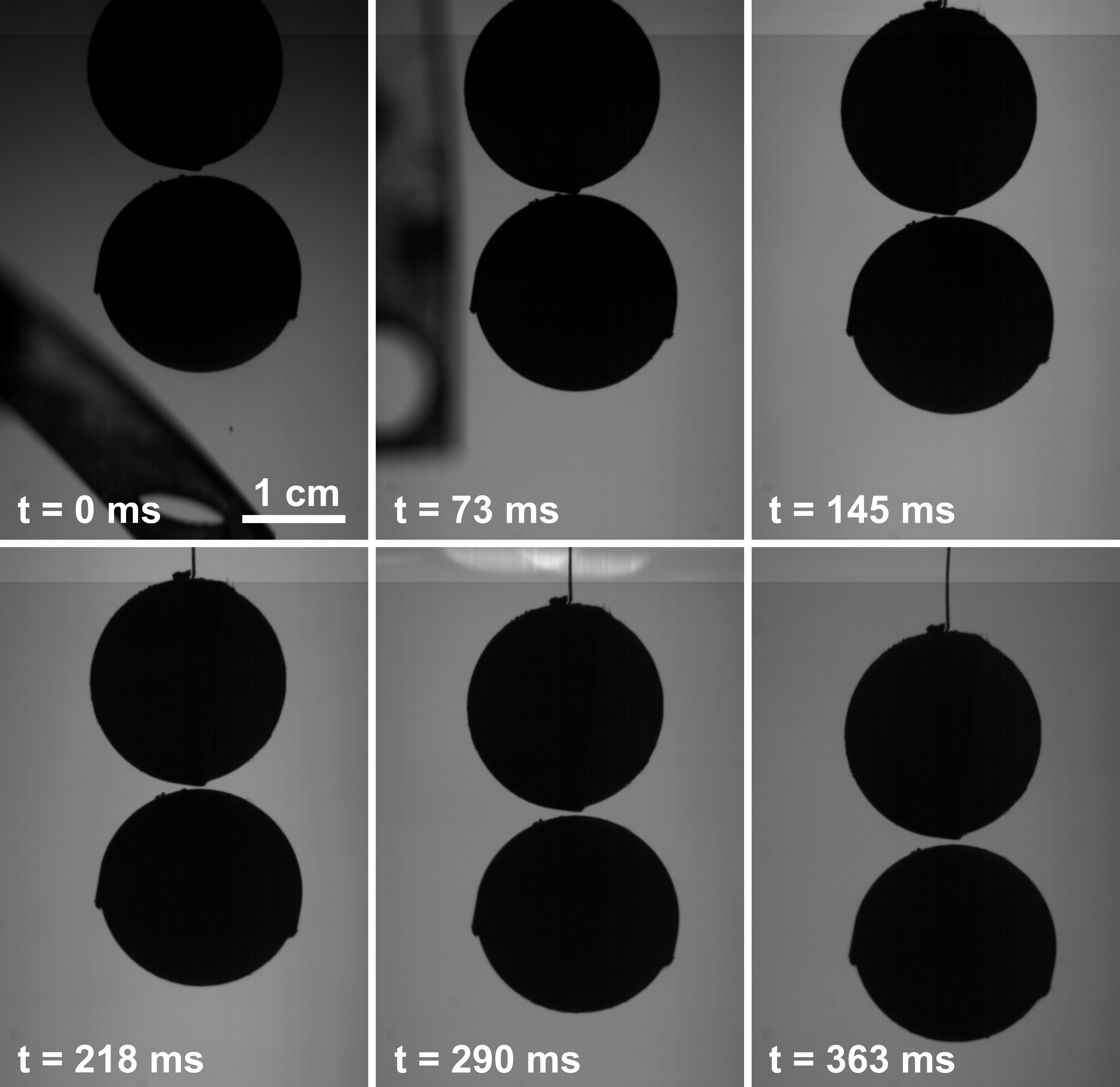}
    \caption{\label{fig:bouncing}A typical sequence of a bouncing collision at an impact velocity of $v=8$~\mms. The coefficient of restitution for this experiment was $\varepsilon=0.29$.}
\end{figure}

\subsection{Bouncing Collisions}\label{sec:BC}
In a bouncing collision the two aggregates approach each other, lose energy in the dissipative contact, and depart with a smaller velocity than before the collision. An appropriate parameter to characterize these collisions is the coefficient of restitution
\begin{equation}
    \varepsilon = \frac{v_\mathrm{after}}{v_\mathrm{before}}.
\end{equation}
The relation $1 - \varepsilon^2$ is a measure for the energy dissipation. Figure \ref{fig:bouncing} shows a typical sequence of a bouncing collision, where the impact velocity was  $v=8$~\mms. The second image is close to the moment of contact and already these raw images at constant time interval allow a good guess for the coefficient of restitution (the exact value was calculated to $\varepsilon = 0.29$).

The main parameters to measure in these collisions are the velocities before and after the collision. To achieve this, we binarized the image sequences of both cameras and convolved them with a disk of the proper sphere diameter to compute the temporally resolved, three dimensional position of the two aggregates. This yielded a precision of 2 pixels, therefore  the quality of the three dimensional distance between the center of the two particles is less than 0.2 mm for a single frame, which becomes better when using a sequence of images. The distance was separately fitted in all three dimensions (for the vertical direction we chose the fast camera) to get a velocity vector before and after the collision. From these, we can calculated the coefficient of restitution $\varepsilon$, the normalized impact parameter $b/R$, and the absolute impact velocity $v$. A rotation of the lower particle was not found in any case. In some cases the upper particle suspended at the embedded nylon string librated, which led to a rotation of the particle in the free fall. To measure the rotation energy and the surface velocity, the position of two points on the surface relative to the center were followed and the surface velocity and the rotation energy calculated.
\begin{figure}[t!]
    \includegraphics[angle=90,width=\columnwidth]{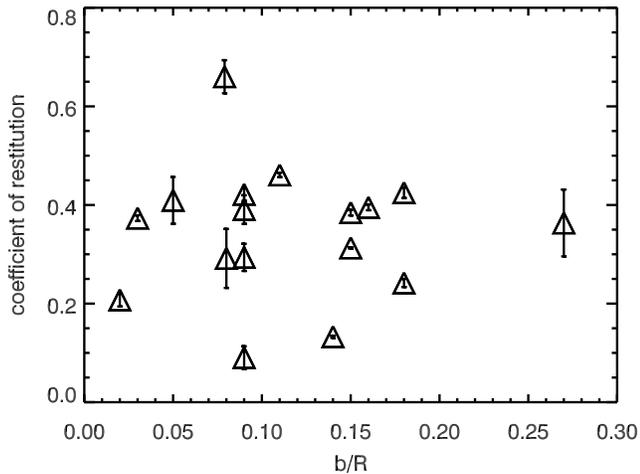}
    \includegraphics[angle=90,width=\columnwidth]{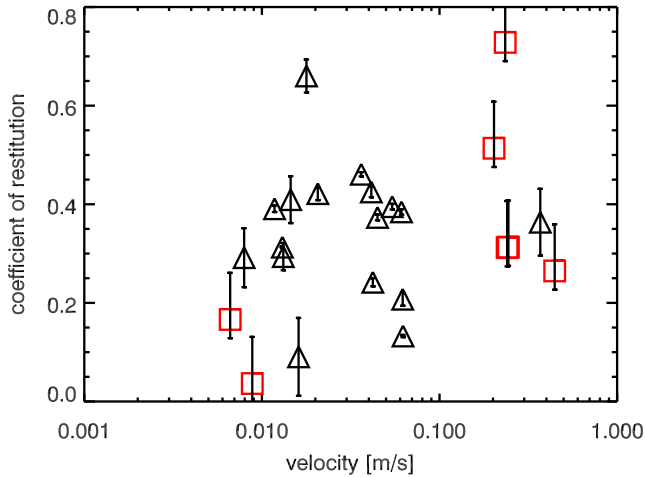}
    \caption{\label{fig:resti}The coefficient of restitution $\varepsilon$ does neither show a correlation with the impact parameter (\emph{top}) nor with the impact velocity (\emph{bottom}). A correlation analysis yields values of $-0.003$ and $0.1$, respectively. Triangles mark all three dimensional data and additionally the 2 two dimensional data are shown (bottom) as squares.}
\end{figure}
That energy was compared with the translational energy and a threshold was set at 10 percent to distinguish between rotating and non rotating impacts.  Below, we will consider only central\ collisions with impact parameters $b/R < 0.3$, and experiments with rotating samples are neglected as well.

Let us first evaluate the dependance of the coefficient of restitution on the normalized impact parameter $b/R$. As shown in Fig. \ref{fig:resti} (top), no strong correlation between these two parameters exists. However, to verify the independence, an analysis for the correlation of these parameters was performed. With this correlation analysis the parameters $b/R$ and $\varepsilon$ could clearly be determined as uncorrelated in the considered parameter range. In Fig. \ref{fig:resti} (bottom) the coefficient of restitution is plotted over the impact velocity. Here, we increased the number of data points by also including 7 collisions where only the high-speed camera was operative, thus where we only have two dimensional information. The collision velocity is dominated by the known vertical component and the coefficient of restitution was corrected by using the average of the horizontal dimension of the photo-camera from the three dimensional cases. This led to a factor of 0.036 that was added to the coefficient of restitution for these collisions. The error bars are as big as the range of the photo-camera component finding in the 3 dimensional collisions. With these 24 collisions, Fig. \ref{fig:resti} (bottom) does not show an obvious correlation between impact velocity and coefficient of restitution and to verify this, we again performed a correlation analysis of these parameters which yielded a correlation coefficient of $0.1$.

\begin{figure}[t]
    \includegraphics[angle=90,width=\columnwidth]{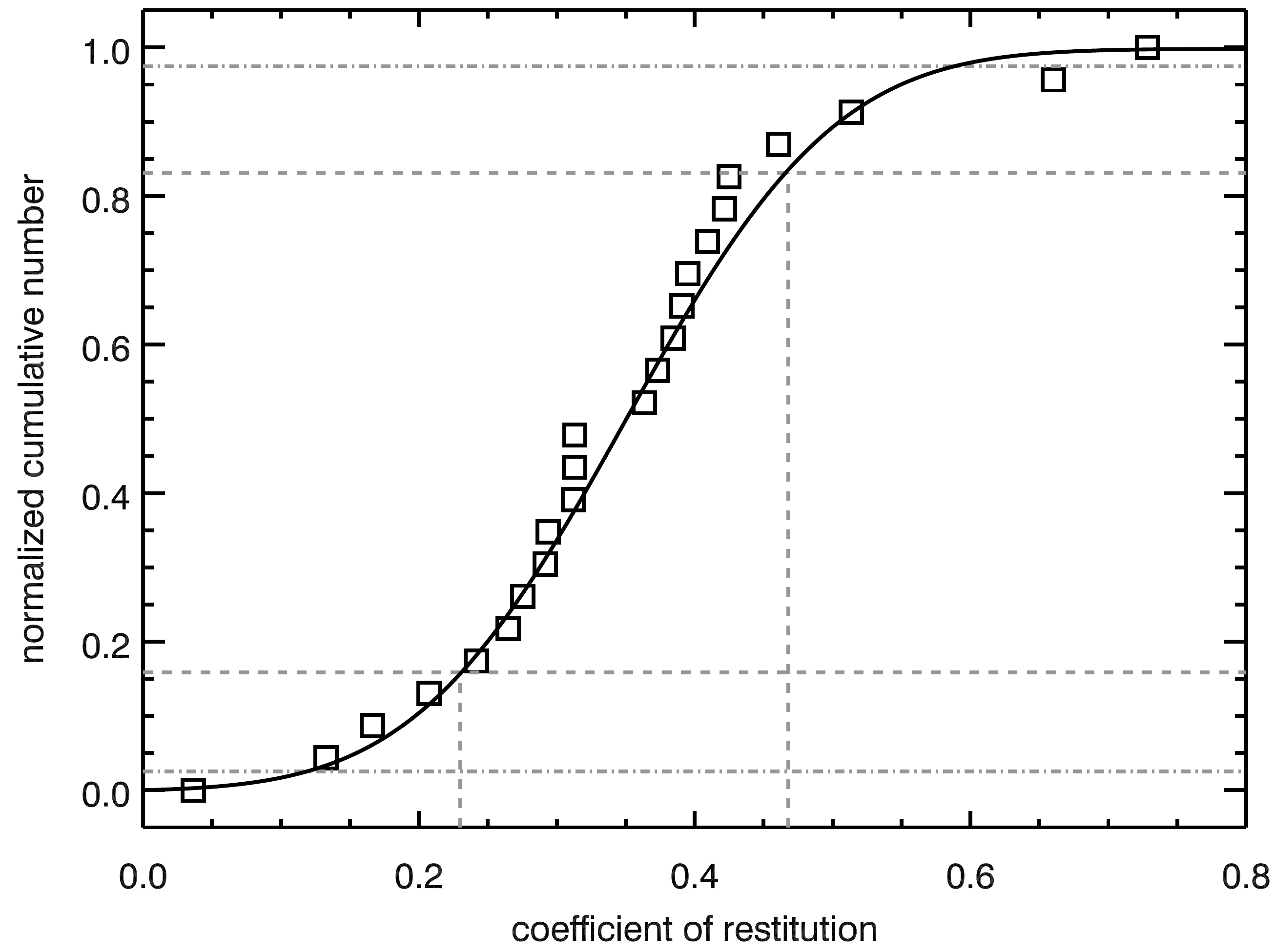}
    \caption{\label{fig:cum_resti}The normalized cumulative number over the coefficient of restitution $\varepsilon$ suggests an integrated Gaussian function what yields to a normal distribution for $\varepsilon$. The average is at $0.35\pm0.12$. The dashed and the dash-dotted lines mark one and two standard deviations, respectively.}
\end{figure}

Conclusively, all impacts can be regarded as cental or near-central collisions and the coefficient of restitution is neither correlated with the impact parameter nor
with the collision velocity in the parameter range considered. We thus plotted the cumulative number fraction over the coefficient of restitution (Fig.
\ref{fig:cum_resti}). The fit is an integrated gaussian function
\begin{equation}
    P(\varepsilon) = \int_0^\varepsilon
    \frac{1}{\sigma_{\varepsilon}\sqrt{2\pi}}\exp\left(-\frac{(\varepsilon'-\overline{\varepsilon})^2}{2\sigma^2_{\varepsilon}}\right)\,\mathrm{d}\varepsilon'
\end{equation}
with an average coefficient of restitution of $\overline{\varepsilon} = 0.35$ and a standard deviation of $\sigma_{\varepsilon}=0.12$. The gray dashed and the dash-dotted lines mark one and two standard deviations, respectively.

\subsection{Fragmenting Collisions}\label{sec:PFC}
Increasing the impact velocity led to aggregate fragmentation. In these cases, the coefficient of restitution could not be calculated, but the strength of fragmentation $\mu$ is the characterizing parameter which we determined. It is defined as $\mu_\mathrm{i} = M_\mathrm{f,i}/M_{\mathrm{0,i}}$ were $M_\mathrm{f,i}$ is the largest fragment and $M_{\mathrm{0,i}}$ is the total mass of the aggregate $i$ (with $i=1$ or 2) before fragmentation. The collision velocity was determined in the same way as in Sect. \ref{sec:BC} and the size of the largest fragment was measured on one or more representative images on the photo camera. Again, these images were binarized and the cross-sectional areas of the fragments were fitted with ellipses of the same expanse. A lower estimate for the fragment volume follows from an ellipsoid where the ellipse is rotated around the long axis and the upper estimate is for rotation around the short axis. Our best estimate is the mean value between those two and the error follows from the values themselves. To calculate the mass of the fragments we assumed that the volume filling factor remained unchanged. The results on the fragmentation strength (i.e. the relation $\mu(v)$) will be discussed in Sect. \ref{sec:discussion}.

\begin{figure}[t]
    \includegraphics[width=\columnwidth]{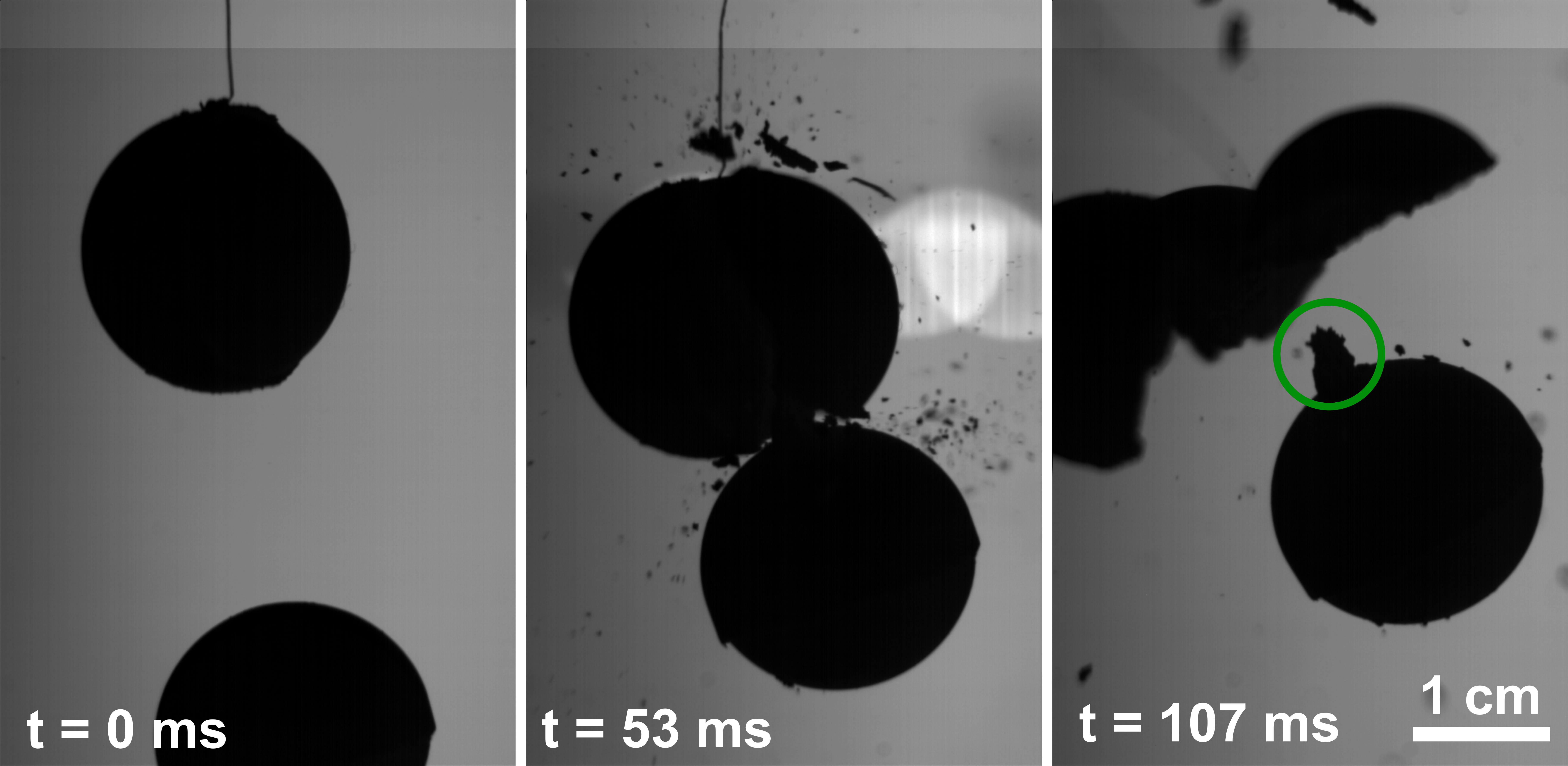}
    \caption{\label{fig:fragmentation}An image sequence of a fragmenting collision at a velocity of 47~\cms. The upper aggregate is destroyed while the lower one gains 0.6 percent of its original volume.}
\end{figure}

The slowest impact velocity where fragmentation occurred was at 24~\cms\ and the fastest bouncing collision was at 37~\cms. In Fig. \ref{fig:fragmentation}, a sequence of a collision at 47~\cms\ is presented. This is a typical collision if the velocity is below 187~\cms, as for these velocities only one aggregate fragmented while the other one was intact and even gained mass. This roughly conically shaped dust pile is marked with a green circle in the last image of Fig. \ref{fig:fragmentation}. We assume this pile is rotational symmetric, we could easily compute its volume: we rotated the image such that the long axis was vertical and measured the width of the pile for each line. Due to the symmetry, each line represented a disk with a height of one pixel and by summing up these disks we got a volume. Again, we assumed that the volume filling factor was unchanged to get the mass of the dust pile. In the example in Fig. \ref{fig:fragmentation} the intact aggregate has grown by 0.57 percent of its own mass. The mass gain for all 6 experiments ranges from 0.05 to 1.1 percent and the velocity dependence will be presented in Sect. \ref{sec:R_ShootingSetup}. There was no relation on whether the upper or lower particle fragmented as these two possibilities exactly split up to the 6 experiments. However, the effect that only one particle was fragmented had a clear limitation at a maximum velocity of about 190~\cms. The only exception was one collision at $v=47$~\cms, in which both aggregates fragmented. For all collisions above 190~\cms\ both particles fragmented.

\begin{figure}[t]
    \includegraphics[width=\columnwidth]{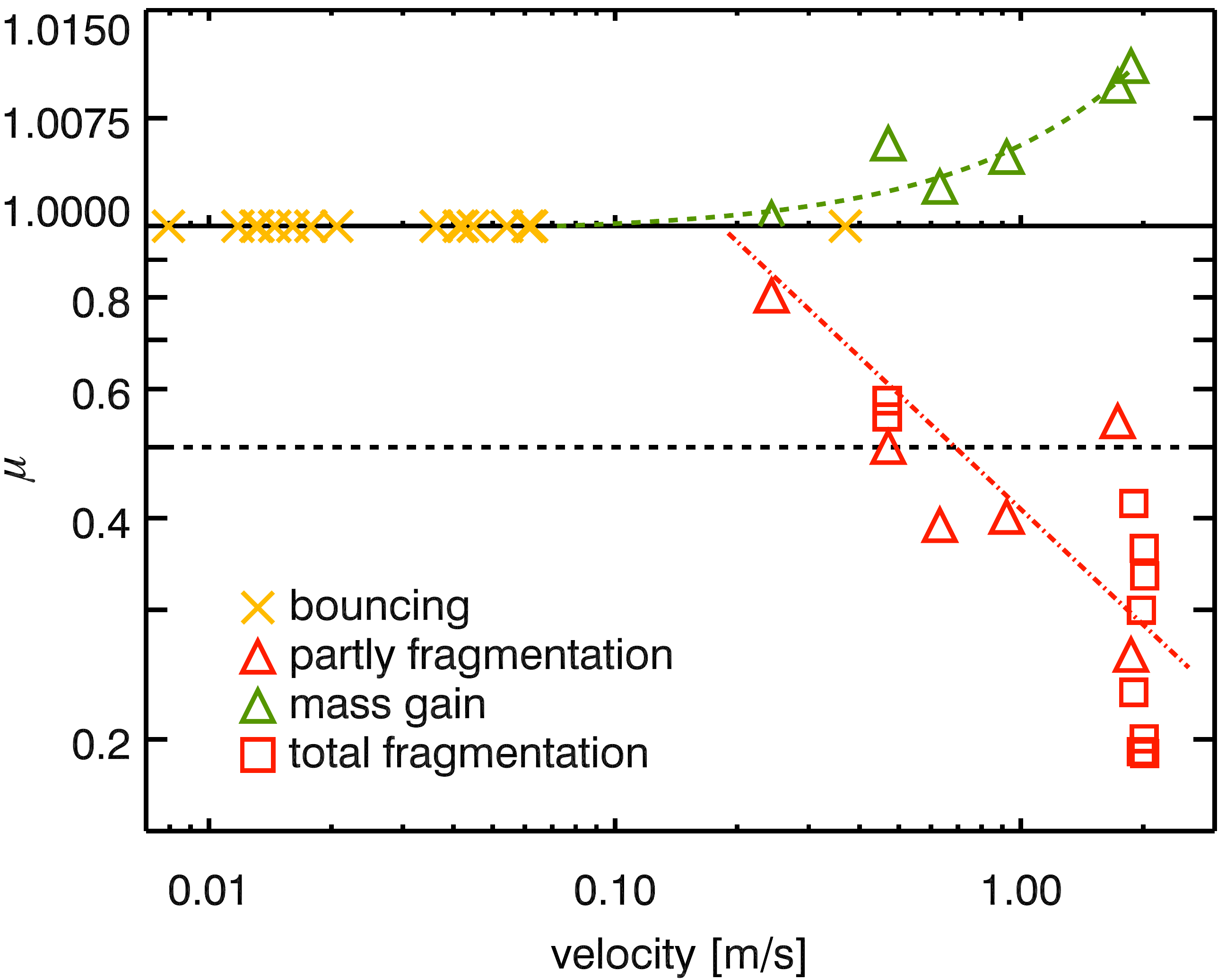}
    \caption{\label{fig:mu}The strength of fragmentation $\mu$ over the impact velocity for all drop-tower collisions. The axis for $\mu>1$ was stretched for better visibility. $\mu$ is plotted for each individual sphere so that we get two data points for each experiment. The values for fragmentation (red) are fitted with a power law (dash-dotted line) and those for mass gain (green) linearly. The black dashed line at $ \mu=0.5$ denotes the limit for catastrophic fragmentation (cf. Sect. \ref{sec:discussion}).}
\end{figure}

Depending on collision velocity, we qualitatively found the three different collisional outcomes that comprise bouncing, fragmentation with mass transfer and complete fragmentation. In Fig. \ref{fig:mu} we present a conclusive overview of the outcomes in the drop-tower experiments. The strength of fragmentation for each individual sphere in a collision $\mu_i$ is plotted as a function of the collision velocity. For the bouncing collision (previous section) the mass remains constant, so that the value for these collisions is unity. Each fragmenting collision yields two $\mu$ values, i.e. one for each aggregate. The disruptive collisions are given by the red squares and those which lead to mass growth yield one negative and one positive $\mu$ value (red and green triangles) for the disrupted and grown aggregate, respectively. We adopt a power law to the values with $\mu<1$, which is given by
\begin{equation}\label{eq:mu_loss}
    \mu^-(v)=\left(\frac{v}{(0.18 \pm 0.04)\ \mathrm{m\,s^{-1}}}\right)^{-0.52 \pm 0.09}
\end{equation}
and shown as a red dash-dotted line. The vertical axis for the collisions with mass gain is linear and stretched and we found that it can be described by a linear relation
\begin{equation}
    \mu^+(v) =   (-4.3 \cdot 10^{-4}+ 6.0 \cdot 10^{-3} \cdot \left(\frac{v}{\mathrm{m\,s^{-1}}}\right))+1 .
\end{equation}
More details on the accretion efficiency will be given in the next section.

\begin{figure}[t]
    \includegraphics[width=\columnwidth]{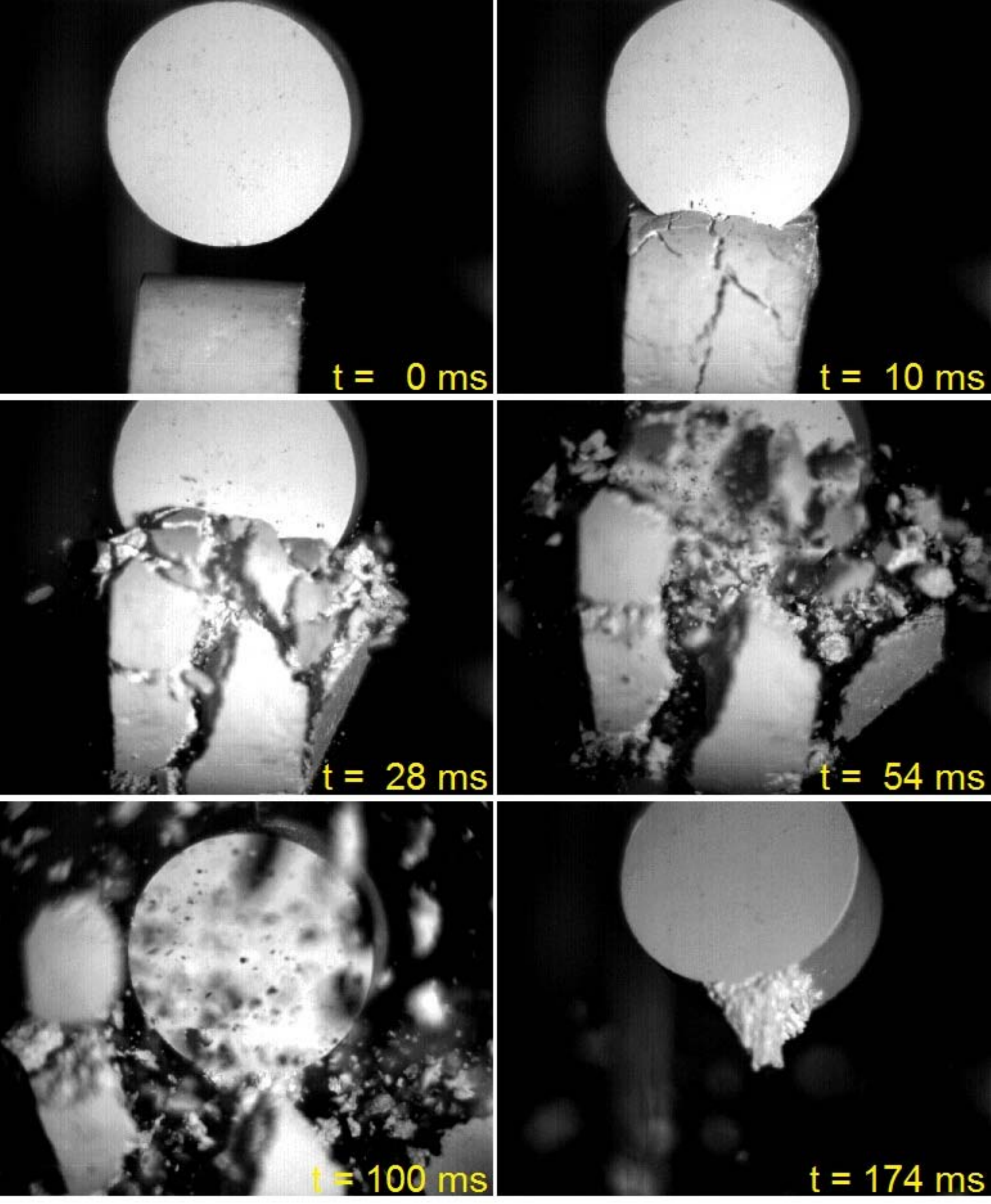}
    \caption{\label{fig:bseq_por1a}An image sequence that illustrates a collision between the cylindrical samples. Here, the collision velocity is roughly 1~\metersecond.}
\end{figure}

\subsection{Accretion Efficiency}\label{sec:R_ShootingSetup}
To further study the accretion efficiency as a function of the porosity difference and collision velocity, we first kept the porosity difference constant at $0.091 \pm 0.014$ for 20 experiments and varied the collision velocity from 0.5 to 1.4~\metersecond. We used the setup shown in Fig. \ref{fig:sole} for this. In a second step, we performed 24 experiments at a constant collision velocity of $(0.93 \pm 0.08)$~\metersecond\ and varied the porosity difference between the aggregates in the range of 0.03 to 0.13. The target aggregate -- the upper aggregate, which was more compact and thus robust -- always gained mass as can be seen in Fig. \ref{fig:bseq_por1a} and after capturing the intact aggregate, we were able to measure the accretion efficiency $e_\mathrm{ac}$, i.e. the mass growth in units of the projectile mass. For a few examples, the mass growth was also deduced from the images with the same approach as presented in the drop-tower experiments. Here the deduction of mass growth from the images are in good agreement with the measured accretion efficiency. For this reason it can be concluded that the volume filling factor of the cone remains unchanged in comparison with the projectile filling factor.

\begin{figure}[t]
    \includegraphics[width=\columnwidth]{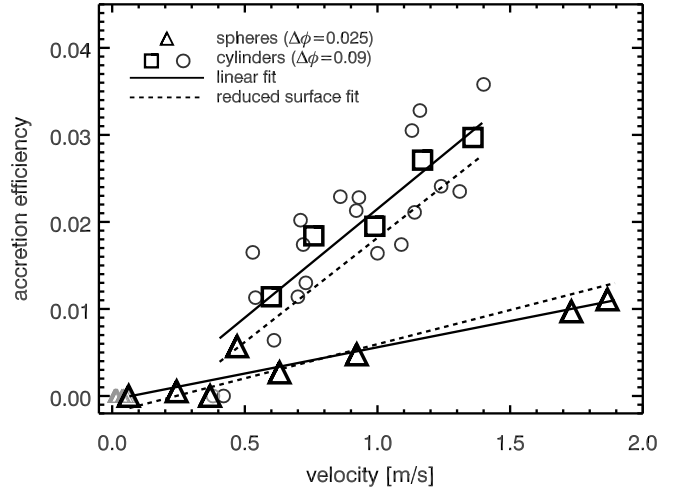}
    \caption{\label{fig:massgrowthvel}Mass gain over collision velocity at constant porosity difference $\Delta\phi=0.091$. Each gray circle represents one experiment, the black line a linear fit to these data points and black squares are averaged values to illustrate the linear trend. Results from the drop-tower are presented as triangles and are also linearly fitted (lower solid line). The dotted lines are the result from the 2 dimensional fit which is discussed in Sect. \ref{sec:discussion}.}
\end{figure}

The mass gain dependence on the collision velocity is presented in Fig. \ref{fig:massgrowthvel}, the gray circles represent single experiments. Below 0.5~\metersecond\ no mass growth but only bouncing collisions without mass transfer were observed (two data points at $e_\mathrm{ac}=0$). Therefore, we can confirm a low-velocity cut off, where collisions are not energetic enough for sufficient inelastic interaction, i.e. to lead to fragmentation. The values above 0.5~\metersecond\ are averaged in steps of 0.2~\metersecond\ up to 1.5~\metersecond, which is shown by the black squares. An increase of the mass gain with collision velocity is clearly visible. Faster collisions lead to a higher mass gain. We assumed a linear increase and fitted a straight line
\begin{equation}\label{eq:Madd1}
	e_\mathrm{ac}\left(v, \Delta\phi=0.091\right) = - 3.5 \cdot 10^{-3} + 2.5 \cdot 10^{-2} \left(\frac{v}{\mathrm{m\,s^{-1}}}\right)
\end{equation}
to the raw data (neglecting the first point at $e_\mathrm{ac}=0$). The black triangles are the results from the experiments in the drop-tower setup presented in the previous section and show a similar trend. The lower values can be explained by a smaller difference in porosity (see below and Sect. \ref{sec:discussion}).

\begin{figure}[t]
    \includegraphics[width=\columnwidth]{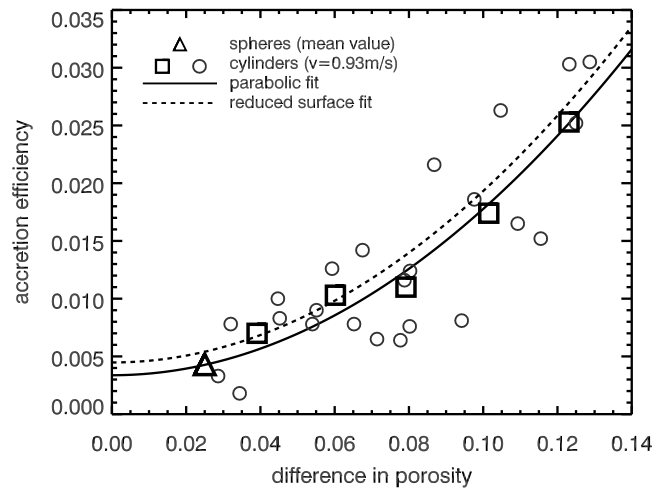}
    \caption{\label{fig:massgrowthpor}Mass gain as a function of porosity difference for a constant collision velocity of 0.93~\metersecond. The gray circles show the raw data, the solid curve is a fit to these data according to Eqs. \ref{eq:parabolic_func} and \ref{eq:exp_func}, these equations are undistinguishable from each other in the parameter range shown, and the black squares represent averaged values. The triangle shows the mean accretion efficiency of all drop tower experiments. The dashed curve is discussed in Sect. \ref{sec:discussion}.}
\end{figure}

The mass gain as a function of the measured porosity difference of colliding aggregates is shown in Fig. \ref{fig:massgrowthpor}, again, the gray circles denote individual experiments. A trend of increasing accretion efficiency with increasing porosity difference is already visible in the raw data and becomes even clearer after averaging in steps of 0.02 in porosity. Without further justification but to give a simple mathematical expression we fitted a parabolic curve to the raw data given as
\begin{equation}\label{eq:parabolic_func}
	e_\mathrm{ac,par}\left(\Delta\phi, v=0.93\ \mathrm{m\,s^{-1}}\right) = 3.4\cdot10^{-3} + 1.44 \cdot \Delta \phi^2\ ,
\end{equation}
which is shown as a solid line in Fig. \ref{fig:massgrowthpor}. Alternatively, an exponential function given as
\begin{equation}\label{eq:exp_func}
	e_\mathrm{ac,exp}\left(\Delta\phi, v=0.93\ \mathrm{m\,s^{-1}}\right) = 4.4\cdot10^{-3} + 1.2\cdot10^{-3} \cdot \exp(23 \cdot \Delta \phi)
\end{equation}
can hardly be distinguished from the previous curve and gives a similar (even slightly better) reduced $\chi^2$. Both curves fit the data in the presented range while they significantly deviate for higher porosity differences: the exponential function gives an accretion efficiency of 0.5 for a porosity difference of 0.26, while for the parabolic function this happens at a porosity difference of 0.59. An extrapolation of this curve is dangerous and not recommended but the consequences would be that the parabolic function predicts a limeted mass transfer even for the highest possible porosity differences while the exponential function predicts a saturation or the onset of a new effect for porosity differences as small as 0.3. This can be tested in future experiments.

\section{Discussion}\label{sec:discussion}
Depending on collision velocity, we qualitatively found three different collisional outcomes that comprise bouncing, fragmentation with mass transfer and complete fragmentation. In Fig. \ref{fig:mu} we presented a conclusive overview of the outcomes in the drop-tower experiments, which give us the velocity range of these effects:\\[0.7em]
\begin{tabular}{p{3cm}cp{4cm}}
    $v < 40$~\cms\       &:& bouncing \\
    $v > 20$~\cms\ and $v<190$~\cms\ &:& partly fragmentation with mass transfer \\
    $v > 190$~\cms\      &:& disruptive fragmentation
\end{tabular}
\\[0.7em]

Bouncing collisions were presented in Sect. \ref{sec:BC} and the main result was the mean coefficient of restitution of $\varepsilon = 0.35\pm0.12$, where the error does not represent the measurement precision but the standard deviation resulting from the natural scatter of the data.
\begin{figure}[t]
    \includegraphics[angle=90,width=\columnwidth]{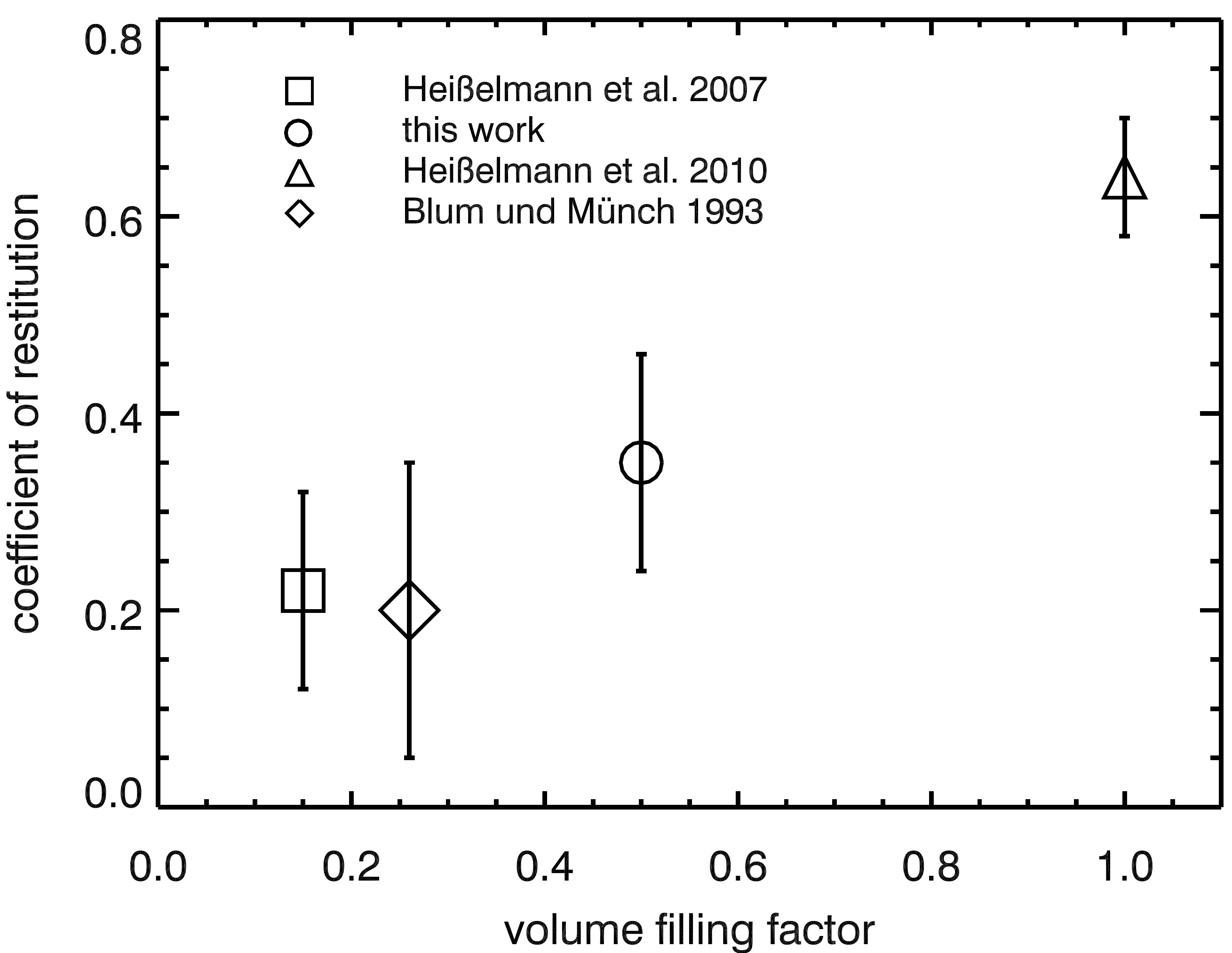}
    \caption{\label{fig:VFF}The coefficient of restitution of various experiments as a function of the volume filling factor. although very clear, this trend should only be regarded as qualitative because of the difference in samples, impact parameter and collision velocity.}
\end{figure}
To compare this results with earlier experiments, we present the coefficient of restitution as a function of the volume filling factor for various experiments (Fig. \ref{fig:VFF}). These experiments are all different: while \citet{HeisselmannEtal:2007} used the same dust material as in our drop-tower experiments and a similar velocity (20~\cms), the aggregates had a cubic shape and the volume filling factor was much lower ($\phi=0.15$). \citet{BlumMuench:1993} performed near central collisions between aggregates of ZrSiO$_4$ dust with a comparable grain size and a volume filling factor of $\phi=0.26$. The fourth data point is from \citet{HeisselmannEtal:2010} who used solid glass beads (still SiO$_2$) with all possible impact parameters in a multiple collision experiment within a range of velocities (0.3 to 5~\cms). In spite of these differences, a clear trend is obvious, which is that porous aggregates behave more inelastic, thus dissipate more energy than compact aggregates or even solid samples. This trend is much stronger than the dependence on the collision velocity, which we could not resolve in Sect. \ref{sec:BC}.

\begin{figure}[t]
    \includegraphics[angle=90,width=\columnwidth]{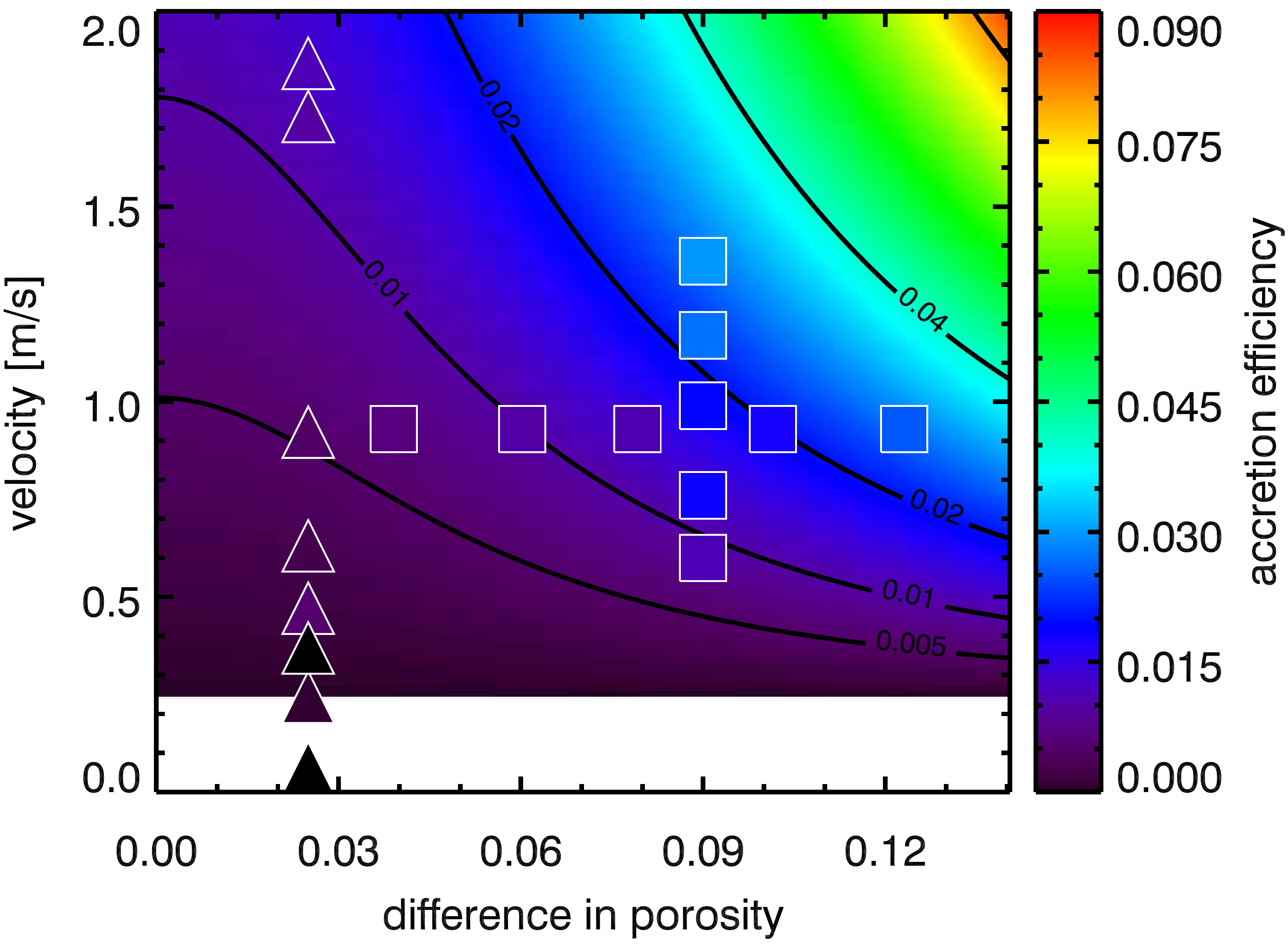}
    \caption{\label{fig:massgrowth_both_param}Two-dimensional fit interpolating the mass accretion as a function of velocity and porosity difference according to Equation (\ref{eq:accretion_efficiency}). The data points represent the measured accretion efficiency in color, for the experiment with cylinders we only show the mean values.}
\end{figure}

For velocities around 20~\cms\ we find that particle fragmentation sets in, which means that only one aggregate is destroyed and transfers mass to the other particle. In  Figs. \ref{fig:massgrowthvel} and \ref{fig:massgrowthpor} we presented the accretion efficiency as a function of collision velocity and porosity difference. From these, we now want to give a full equation $e_\mathrm{ac}(v, \Delta\phi)$. We cannot simply multiply Equations (\ref{eq:Madd1}) and (\ref{eq:parabolic_func}) (or (\ref{eq:exp_func}), respectively) because these do not exactly match at $v=0.93$~\metersecond\ and $\Delta\phi = 0.091$. We therefore fitted the raw data points from the experiments with the cylindrical samples with the function
\begin{eqnarray}\label{eq:accretion_efficiency}
	e_\mathrm{ac}(v, \Delta\phi) &=& \left(a + b \cdot \Delta \phi^2\right) \cdot \left(c+d \cdot v\right)\ .
\end{eqnarray}
A special search technique like the downhill simplex algorithm was used to obtain the parameters. The calculated parameters are $a=0.26$, $b=85$, $c=-6.1 \cdot 10^{-3}$ and $d=2.5\cdot 10^{-2}\ \mathrm{s\,m^{-1}}$. Equation (\ref{eq:accretion_efficiency}) can again be reduced to the constant porosity difference and constant velocity for which the series of experiments were conducted, which is presented in  Figs. \ref{fig:massgrowthvel} and \ref{fig:massgrowthpor}\ as the dashed line. Those are still in the range of the errors and differ only slightly from the original functions. The combination of both parameters is presented in the contour plot of Fig. \ref{fig:massgrowth_both_param}. The color values correspond to Equation (\ref{eq:accretion_efficiency}) and the color of the square symbols represents the value of the averaged data (squares in Figs. \ref{fig:massgrowthvel} and \ref{fig:massgrowthpor}). The white region at low velocities ($v\lesssim25$~\cms) is where Equation (\ref{eq:accretion_efficiency}) mathematically predicts a negative mass transfer, while we rather expect bouncing collisions. This is supported by the black triangles, representing bouncing collisions without mass transfer.\ Due to the difference in aggregate geometry and grain shape, the results from the drop-tower experiments were not utilized to compute the fit in Equation (\ref{eq:accretion_efficiency}) but they are nonetheless very well represented by this curve (see triangles in Fig. \ref{fig:massgrowth_both_param}). If we again reduce it to the constant porosity difference $\Delta\phi=0.025$, we find in Fig. \ref{fig:massgrowthvel} that the dashed line is very close to the solid line. This agreement without further assumption assures us that the experiments are highly comparable in spite of the discussed differences. We therefore regard Equation (\ref{eq:accretion_efficiency}) as a good description for the mass transfer in a disruptive collision between two aggregates with different strength. The validity range of this equation is given by $20\ \mathrm{cm\,s^{-1}} < v < 190\ \mathrm{cm\,s^{-1}}$ in velocity. For the porosity, we cannot make a definite prediction for $\Delta\phi>0.13$, which is the maximum value that was studied here.

\begin{figure}[t]
    \includegraphics[angle=180,width=\columnwidth]{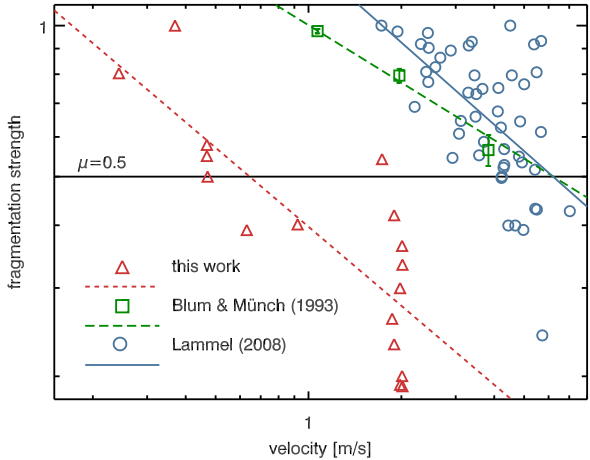}
    \caption{\label{fig:mu_all}Fragmentation strength as a function of impact velocity. The triangles, squares, and circles show the data from Fig. \ref{fig:mu}, from \citet{BlumMuench:1993}, and \citet{Lammel:2008}, respectively. The dashed and solid lines are power-law fits to these data (see Eq. \ref{eq:mu_gen}).}
\end{figure}

The velocity of 20~\cms\ where the fragmentation sets in is significantly lower than expected from experiments with millimeter-sized aggregates \citep{GuettlerEtal:2010}. We therefore expect a size dependence of the threshold velocity and compare our results with data of \citet{BlumMuench:1993} and \citet{Lammel:2008} who measured the strength of fragmentation for a variable velocity as presented here in Sect. \ref{sec:PFC}. We consider central collisions and take the mean values from Fig. 7 in \citet{BlumMuench:1993} as well as raw data of \citet{Lammel:2008}. This is presented in Fig \ref{fig:mu_all} where the red triangles represent our data from Sect. \ref{sec:PFC}, the green squares are the values of \citet[mean values, thus including errors]{BlumMuench:1993} and the blue circles represent the data of \citet{Lammel:2008}. From this we get equations similar to Eq. (\ref{eq:mu_loss}), which relate the fragmentation strength to the collision velocity between two dust aggregates. In its general form, this reads
\begin{equation}\label{eq:mu_gen}
    \mu = \left( \frac{v}{v_0} \right)^{-\alpha}\, .
\end{equation}
Our data, presented in Fig. \ref{fig:mu}, yield $v_0 = (0.17^{+0.07}_{-0.05})$~\metersecond\ and $\alpha = 0.52 \pm 0.10$ for collisions between equal-sized dust aggregates with masses of $m = 4.1$~g and radii of $s = 1$~cm (short dashed red line in Fig. \ref{fig:mu_all}). For collisions between mm-sized dust aggregates, i.e. for the data of \citet{BlumMuench:1993} (using aggregates with $m = 4$~mg and $s = 0.9$~mm consisting of $\rm \mu m$-sized $\rm ZrSiO_4$), we get $v_0 = (1.01 \pm 0.03)$~\metersecond\ and $\alpha = 0.38 \pm 0.04$ (long dashed, green line). For the data of \citet{Lammel:2008} (using aggregates with $m = 1$~mg and $s = 0.7$~mm consisting of $\rm \mu m$-sized $\rm SiO_2$; $s$ is the half edge length of the cubic aggregates here), the parameters are $v_0 = (1.73^{+0.63}_{-0.46})$~\metersecond\ and $\alpha = 0.54 \pm 0.11$ (solid blue line). These relations between fragmentation strength and velocity can be used to derive the velocity for the onset of fragmentation, i.e. $v_{1.0} = v(\mu =1.0) = v_0$ and for the case that the largest fragment possesses half the mass of the original projectile aggregate, i.e. $v_{0.5} = v(\mu =0.5) = v_0 \cdot 0.5^{-1/\alpha}$.
\begin{figure}[t]
    \includegraphics[angle=180,width=\columnwidth]{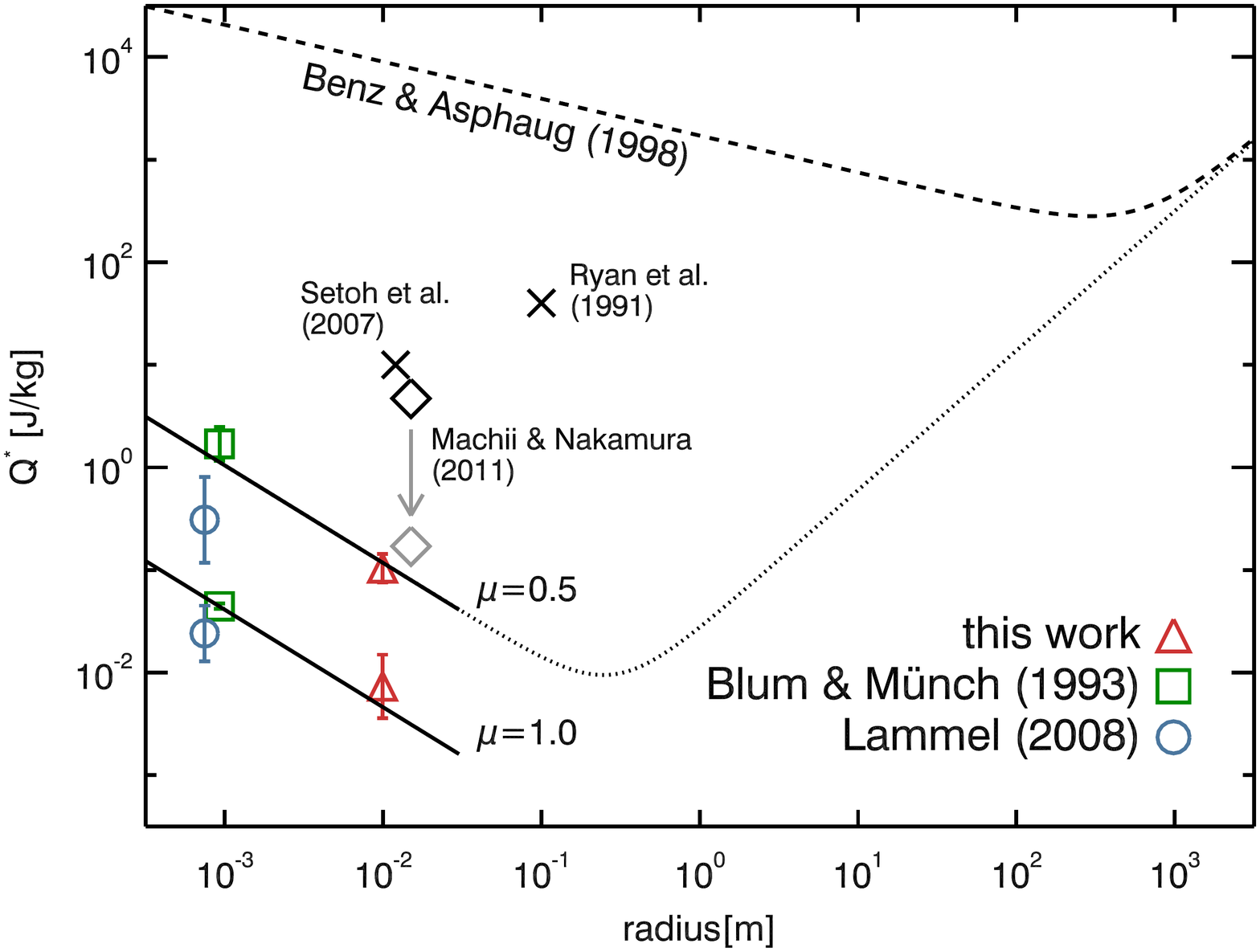}
    \caption{\label{fig:qstar} The critical fragmentation strength $Q^{\ast}$ as a function of dust-aggregate radius for our data (triangles) as well as for the data of \citet{BlumMuench:1993} (squares) and \citet{Lammel:2008} (circles). A least-squares power law fit after Eq. \ref{eq:qstar} is shown as solid line for the two cases $\mu = 1.0$ and $\mu = 0.5$. Also shown are the $Q^{\ast}$ data of \citet{SetohEtal:2007} and \citet{RyanEtal:1991} (crosses) as well as the curve by \citet{BenzAsphaug:1999} (dashed line). The dotted line is an extrapolation of our data according to the model by \citet{BenzAsphaug:1999}. }
\end{figure}
Applying this to all the data in Fig. \ref{fig:mu_all}, we get $v_{0.5} = 0.64$~\metersecond\ for our data, $v_{0.5} = 6.2$~\metersecond\ for the data of \citet{BlumMuench:1993} as well as for the data of \citet{Lammel:2008}; the velocity for the onset of fragmentation is already given by the fit parameters $v_0$. These velocities can be converted into specific energies, assuming equal-mass collision partners, by the relation $Q_x = \frac{1}{4} v_x^2$, with $x=1.0$ or $x=0.5$. We want to bring forward a relation for the critical fragmentation energy depending on aggregate size but have to be careful in comparing these data as the volume filling factors significantly differ from 0.15 \citep{Lammel:2008} to 0.5 (this work).

The porosity has an effect on the strength of the aggregates as well as on the transmission of impact stress waves. The tensile strength for these aggregates is given by
\begin{equation}
    T(\phi) = -(10^{2.8+1.48\cdot\phi})
\end{equation}
\citep[][their Eq. 8]{GuettlerEtal:2009} and for the relation between aggregate strength and fragmentation energy we consider the results of \citet{SetohEtal:2010} and \citet{MachiiNakamura:2011} and take $Q \propto T^{3/4}$. The effect of poor shock transmission in porous media was studied by \citet{LoveEtal:1993} who found a porosity dependence of $Q \propto \phi^{-3.6}$ if the strength of the bodies remains constant for changing porosity. In their numerical simulation on the disruption of larger bodies, also \citet{JutziEtal:2010} found that porous targets are more difficult to disrupt. Thus, following \citet{LoveEtal:1993}, we apply a correction as
\begin{equation}
    Q_x(\phi=0.5) = Q_x(\phi) \frac{0.5^{-3.6}10^{1.11\cdot0.5}}{\phi^{-3.6}10^{1.11\cdot\phi}},
\end{equation}
where $Q_x(\phi)$ is the critical energy for the individual experiments as calculated above and $Q_x(\phi=0.5)$ is the critical energy corrected to a volume filling factor of $\phi=0.5$ as used in our experiments. With this, we arrive at Fig. \ref{fig:qstar}, where the critical fragmentation energies $Q_{0.5}$ and $Q_{1.0}$ for the three experiments are presented as a function of aggregate size. A power-law fit to these data of the form
\begin{equation}\label{eq:qstar}
    Q_x^{\ast} = \left( \frac{s}{s_x} \right)^{-\beta} \, \rm J~kg^{-1}
\end{equation}
yields $\beta = 0.95 \pm 0.38$, $s_{0.5} = (1.0^{+0.3}_{-0.2}) \cdot 10^{-3}$~m, and $s_{1.0} = (3.4^{+2.8}_{-1.2}) \cdot 10^{-5}$~m. We have to note that the slope should be regarded as an estimate and it is desirable to have another experiment to measure the fragmentation energy for different sized samples but with the same strength and porosity. Accordant studies are currently on the way.
Our slope is slightly steeper than in the material-strength regime presented by \citet{BenzAsphaug:1999} (see Fig. \ref{fig:qstar}), who studied collisions between rocky bodies in computer simulations. Our firm absolute values for $Q_{0.5}^{\ast}$ are several orders of magnitude smaller than those experimentally measured by \citet{SetohEtal:2007} and \citet{RyanEtal:1991} (see Fig. \ref{fig:qstar}), who used sintered glass beads and weakly glued gravel, respectively. \citet{MachiiNakamura:2011} also performed experiments with sintered glass beads (black diamond in Fig. \ref{fig:qstar}) and found a dynamic strength in the same order of magnitude as \citet{SetohEtal:2007} and \citet{RyanEtal:1991}. However, they also measured the static tensile strength of the material ($3.5 \cdot 10^5$~Pa) and if we apply a similar scaling as above, we arrive at the gray diamond which is in good agreement with our results.  Although not perfectly applicable for our samples, we use the model by \citet{BenzAsphaug:1999} for the strength in the gravitational regime. Actually, in a recent study of \citet{JutziEtal:2010} the slopes in the gravity regime (cf. their Table 3) are mostly comparable to those of \citet{BenzAsphaug:1999}, also if they use porous bodies, so that we take 

\begin{equation}\label{eq:gravity}
    Q_{0.5}^{\ast} = \left\{ \left( \frac{s}{1\ \mathrm{mm}} \right)^{-0.95} + \left(\frac{s}{15\ \mathrm{m}}\right)^{1.36} \right\}\ \rm J~kg^{-1}\\
\end{equation}
\\
\\
illustrated by the dotted line in Fig. \ref{fig:qstar}. The first term is simply given by Eq. \ref{eq:qstar} while the second term is taken from the gravity regime in \citet[cf. their Eq. 6 and Table 3]{BenzAsphaug:1999} assuming a bulk density of 1000~kg~m$^{-3}$. Thus, our measurements suggest that the weakest bodies in PPDs have radii of the order of 30 cm. Their strength should be around $Q^{\ast} = 10^{-2}\ \mathrm{J\,kg^{-1}}$, which corresponds to impact velocities between equal-sized particles of about 0.2~\metersecond.\ A comparison to the expected collision velocities in different PPD models \citep[see Fig. 8 in][]{WeidlingEtal:2009} shows that the collision velocities for decimeter-sized aggregates exceed this critical fragmentation velocity in both, the minimum-mass solar nebula model \citep{Weidenschilling:1977b} as well as in the low-density model by \citet{AndrewsWilliams:2007}. The collision velocities in the high-density model by \citet{Desch:2007} are comparable to our estimated fragmentation velocity for decimeter-sized dust aggregates. Thus, dust aggregates of decimeter size can only barely survive mutual collisions if the impact velocities are very moderate. Alternatively, the strength of aggregates against fragmentation increases if the aggregates experience any solidification process, e.g. by sintering (see data by \citet{SetohEtal:2007} and \citet{RyanEtal:1991} in Fig. \ref{fig:qstar}).

\subsection*{Acknowledgements}
We are grateful to the Deutsches Zentrum für Luft- und Raumfahrt (DLR) for funding the experiments in the Braunschweig laboratory under grants 50WM0636 and 50WM0936. C.G. and T.M. want to thank the Deutsche Forschungsgemeinschaft for funding the Forschergruppe 759 ``The Formation of Planets: The Critical First Growth Phase'' (grants Bl 298/14-1 and Wu 321/5-2), which also funded the impact experiments in the Duisburg laboratory. Supporting scanning electron microscopy measurements were performed at the Max Planck Institute for Polymer Research in Mainz. We thank Hans-J\"urgen Butt for providing us with this possibility.

\bibliographystyle{aa}
{\bibliography{literatur}}

\begin{thebibliography}{22}
\expandafter\ifx\csname natexlab\endcsname\relax\def\natexlab#1{#1}\fi

\bibitem[{{Andrews} \& {Williams}(2007)}]{AndrewsWilliams:2007}
{Andrews}, S.~M. \& {Williams}, J.~P. 2007, \apj, 659, 705

\bibitem[{{Benz} \& {Asphaug}(1999)}]{BenzAsphaug:1999}
{Benz}, W. \& {Asphaug}, E. 1999, Icarus, 142, 5

\bibitem[{{Blum} \& {M{\"u}nch}(1993)}]{BlumMuench:1993}
{Blum}, J. \& {M{\"u}nch}, M. 1993, Icarus, 106, 151

\bibitem[{{Blum} \& {Schr{\"a}pler}(2004)}]{BlumSchraepler:2004}
{Blum}, J. \& {Schr{\"a}pler}, R. 2004, \prl, 93, 115503

\bibitem[{{Blum} \& {Wurm}(2008)}]{BlumWurm:2008}
{Blum}, J. \& {Wurm}, G. 2008, \araa, 46, 21

\bibitem[{{Desch}(2007)}]{Desch:2007}
{Desch}, S.~J. 2007, \apj, 671, 878

\bibitem[{{G{\"u}ttler} {et~al.}(2010){G{\"u}ttler}, {Blum}, {Zsom}, {Ormel},
  \& {Dullemond}}]{GuettlerEtal:2010}
{G{\"u}ttler}, C., {Blum}, J., {Zsom}, A., {Ormel}, C.~W., \& {Dullemond},
  C.~P. 2010, \aap, 513, A56

\bibitem[{{G{\"u}ttler} {et~al.}(2009){G{\"u}ttler}, {Krause}, {Geretshauser},
  {Speith}, \& {Blum}}]{GuettlerEtal:2009}
{G{\"u}ttler}, C., {Krause}, M., {Geretshauser}, R.~J., {Speith}, R., \&
  {Blum}, J. 2009, \apj, 701, 130

\bibitem[{{Heim} {et~al.}(1999){Heim}, {Blum}, {Preuss}, \&
  {Butt}}]{HeimEtal:1999}
{Heim}, L.-O., {Blum}, J., {Preuss}, M., \& {Butt}, H.-J. 1999, \prl, 83, 3328

\bibitem[{{Hei{\ss}elmann} {et~al.}(2010){Hei{\ss}elmann}, {Blum}, {Fraser}, \&
  {Wolling}}]{HeisselmannEtal:2010}
{Hei{\ss}elmann}, D., {Blum}, J., {Fraser}, H.~J., \& {Wolling}, K. 2010,
  Icarus, 206, 424

\bibitem[{{Hei{\ss}elmann} {et~al.}(2007){Hei{\ss}elmann}, {Fraser}, \&
  {Blum}}]{HeisselmannEtal:2007}
{Hei{\ss}elmann}, D., {Fraser}, H., \& {Blum}, J. 2007, in Proceedings of the
  58th International Astronautical Congress 2007, {IAC-07-A2.1.02}

\bibitem[{{Jutzi} {et~al.}(2010){Jutzi}, {Michel}, {Benz}, \&
  {Richardson}}]{JutziEtal:2010}
{Jutzi}, M., {Michel}, P., {Benz}, W., \& {Richardson}, D.~C. 2010, Icarus,
  207, 54

\bibitem[{{Lammel}(2008)}]{Lammel:2008}
{Lammel}, C. 2008, {Experimentelle Untersuchungen zur Fragmentation von
  Staubagglomeraten im Zweiteilchensto{\ss} bei mittleren Geschwindigkeiten},
  Bachelor's thesis, Technische Universit\"at Carolo Wilhelmina zu Braunschweig

\bibitem[{{Love} {et~al.}(1993){Love}, {H{\"o}rz}, \&
  {Brownlee}}]{LoveEtal:1993}
{Love}, S.~G., {H{\"o}rz}, F., \& {Brownlee}, D.~E. 1993, Icarus, 105, 216

\bibitem[{{Machii} \& {Nakamura}(2011)}]{MachiiNakamura:2011}
{Machii}, N. \& {Nakamura}, A.~M. 2011, Icarus, 211, 885

\bibitem[{{Ryan} {et~al.}(1991){Ryan}, {Hartmann}, \& {Davis}}]{RyanEtal:1991}
{Ryan}, E.~V., {Hartmann}, W.~K., \& {Davis}, D.~R. 1991, Icarus, 94, 283

\bibitem[{{Setoh} {et~al.}(2007){Setoh}, {Hiraoka}, {Nakamura}, {Hirata}, \&
  {Arakawa}}]{SetohEtal:2007}
{Setoh}, M., {Hiraoka}, K., {Nakamura}, A.~M., {Hirata}, N., \& {Arakawa}, M.
  2007, Advances in Space Research, 40, 252

\bibitem[{{Setoh} {et~al.}(2010){Setoh}, {Nakamura}, {Michel}, {Hiraoka},
  {Yamashita}, {Hasegawa}, {Onose}, \& {Okudaira}}]{SetohEtal:2010}
{Setoh}, M., {Nakamura}, A.~M., {Michel}, P., {et~al.} 2010, Icarus, 205, 702

\bibitem[{{Weidenschilling}(1977)}]{Weidenschilling:1977b}
{Weidenschilling}, S.~J. 1977, \apss, 51, 153

\bibitem[{{Weidenschilling} \& {Cuzzi}(1993)}]{WeidenschillingCuzzi:1993}
{Weidenschilling}, S.~J. \& {Cuzzi}, J.~N. 1993, in Protostars and Planets III,
  ed. E.~H. {Levy} \& J.~I. {Lunine}, 1031--1060

\bibitem[{{Weidling} {et~al.}(2009){Weidling}, {G{\"u}ttler}, {Blum}, \&
  {Brauer}}]{WeidlingEtal:2009}
{Weidling}, R., {G{\"u}ttler}, C., {Blum}, J., \& {Brauer}, F. 2009, \apj, 696,
  2036

\bibitem[{{Zsom} {et~al.}(2010){Zsom}, {Ormel}, {G{\"u}ttler}, {Blum}, \&
  {Dullemond}}]{ZsomEtal:2010a}
{Zsom}, A., {Ormel}, C.~W., {G{\"u}ttler}, C., {Blum}, J., \& {Dullemond},
  C.~P. 2010, \aap, 513, A57

\end{thebibliography}

\end{document}